\documentclass{ptephy_v1}

\preprintnumber{XXXX-XXXX} 
\usepackage{hyperref}

\usepackage{graphicx}
\usepackage{mathptmx}      
\usepackage{flushend}

\def\eqn#1{\begin{eqnarray}#1\end{eqnarray}}
\def\={&=&}
\def\l{\left}
\def\r{\right}
\def\F{{\cal F}}
\def\nn{\nonumber \\}

\begin{document}

\title{Boundary Conditions for Constraint Systems in Variational Principle}


\author[1,2]{Keisuke \sc{Izumi}}
\affil[1]{Kobayashi-Maskawa Institute, Nagoya University, Nagoya 464-8602, Japan}
\affil[2]{Department of Mathematics, Nagoya University, Nagoya 464-8602, Japan \email{izumi@math.nagoya-u.ac.jp}}

\author[3]{Keigo \sc{Shimada}}
\affil[3]{Department of Physics, Tokyo Institute of Technology, 2-12-1 Ookayama, Meguro-ku, Tokyo 152-8551, Japan \email{k-appliedphysics@akane.waseda.jp}}

\author[4]{Kyosuke \sc{Tomonari}}
\affil[4]{Department of Physics, Tokyo Institute of Technology, 2-12-1 Ookayama, Meguro-ku, Tokyo 152-8551, Japan \email{ktomonari.phys@gmail.com}}

\author[5,6]{Masahide \sc{Yamaguchi}} 
\affil[5]{Cosmology, Gravity and Astroparticle Physics Group, Center for Theoretical Physics of the Universe, Institute for Basic Science, Daejeon 34126, Korea}
\affil[6]{Department of Physics, Tokyo Institute of Technology, 2-12-1 Ookayama, Meguro-ku, Tokyo 152-8551, Japan
\email{gucci@ibs.re.kr}}


\begin{abstract}%
We show the well-posed variational principle in constraint systems. In a naive procedure of the variational principle with constraints, the proper number of boundary conditions does not match with that of physical degrees of freedom dynamical variables, which implies that, even in theories with up to first order derivatives, the minimal (or extremal) of the action with the boundary terms is not a solution of equation of motion in the Dirac procedure of constrained systems. We propose specific and concrete steps to solve this problem. These steps utilize the Hamilton formalism, which allows us to separate the physical degrees of freedom from the constraints. It reveals the physical degrees of freedom which is necessary to be fixed on boundaries, and also enables us to specify the variables to be fixed and the surface terms.
\end{abstract}

\subjectindex{A00, E00, E03}

\maketitle
\section{Introduction}
Constrained systems appear in many areas of physics, such as gauge theories and theories of gravity. For details see the textbook e.g. \cite{Henneaux:1992ig}. Most of theories are described through Lagrangians and their equations of motion are obtained through the variational principle, where boundary conditions are required to be imposed. Since the imposition of the boundary conditions in the variational principle implies that the initial and / or final conditions are given in the equations of motion, the number of imposed boundary condition in the variational principle should match with the necessary and sufficient number of boundary condition to obtain a solution from the equations of motion. 

Despite its importance, the boundary conditions for the variational principle are only vaguely explored in theories with constraints. In the paper~\cite{Dyer:2008hb}, the action
\eqn{
S=\int^{t_2}_{t_1} dt \frac12 \l(\sqrt{m}\dot q_1\pm \sqrt{M}\dot q_2\r)^2-\frac12M\omega^2 q^2_2
\label{eqsimpleac}
}
was introduced as an example. A possible set of naive boundary conditions in the variational principle here might be $\delta q_{1}=\delta q_{2}=0$ at the initial and final points, $t=t_{1},t_{2}$. However, we should impose only one boundary condition at each boundary in the equation of motion, otherwise the system becomes over-constrained. This implies that also in the variational principle to obtain the equation of motion more than one boundary condition should not be imposed at each boundary, that is, imposing boundary conditions $\delta q_{1}=\delta q_{2}=0$ in the variation principle is inconsistent with the Dirac procedure. In fact, since the action \eqref{eqsimpleac} can be obtained from a point particle action 
\eqn{
S=\int^{t_2}_{t_1} dt \frac12 \dot q^{2}_3-\frac12M\omega^2 q^2_2
} 
by changing the variables $q_3 \equiv \sqrt{m} q_1\pm \sqrt{M} q_2$, the authors concluded that the correct boundary conditions are $\delta q_3(t_1) = \delta q_3(t_2) =0$. The reasonable boundary conditions can be imposed in this example because a 'seed' action, from which the constraint systems originated, is obvious. For general cases, however, the {\it systematical} approach has not yet been explored. We emphasize that, only after boundary conditions are adequately fixed, necessary boundary (counter) terms can be elucidated to make the variational principle well-posed. This is the main topic that we address in this paper. For example, many gravity theories often introduce counter terms such as Gibbons-Hawking-York counter term by simply imposing the Dirichlet boundary conditions. However, it may encounter the same difficulties since these theories are also degenerate systems. Actually, some controversies have already arisen, e.g., whether Gibbons-Hawking-York type counter term is necessary in Palatini theories of gravity ~\cite{Brown:1992br,Gomez:2021roj,Saez-ChillonGomez:2020afj}, even though these theories have no second derivative in the action function. In order to address this kind of question adequately, we first should reconsider {\it{adequate boundary conditions}} in degenerate systems\footnote{In this paper, we consider boundary conditions originated from the degeneracy / singularity of Lagrangian, not from the continuum limit of boundary in field theories \cite{Jabbari1999}.} which is consistent with the variational principle. This is what we would like to address in this paper.

The construction of this paper is as follows. We first review the boundary conditions for non-constrained / non-degenerate systems. We show that, for such systems, non-degeneracy directly implies that the number of necessary boundary conditions precisely coincides with the number of variables in the theory. Then in \S\ref{sec:constrained} we first point out the mismatch of the number of boundary conditions and the number of variables for degenerate systems. For the consistency in the variational principle, only the same number as the physical degrees of freedom, which is less than the number of variables in constrained systems, can be imposed. To construct the well-posed variational principle, we then outline systematic steps to obtain the correct number and form of boundary conditions. In \S\ref{sec:examples}, we demonstrate the proposed steps in some examples and show that the well-posed boundary conditions in the variational principle are obtained. We also see in \S\ref{sec:gene} that our procedures work well in general theory to some extent and explicit formulae for variables to be fixed and surface terms are given. Finally, in \S\ref{sec:con} we conclude our findings and expand on possible application to field theories.

\section{Recap of boundary terms for unconstrained systems}
\label{sec:nonconstrained}
First of all, let us review the boundary terms for unconstrained systems. Consider the following action of an $N$ point particles system
\eqn{
S =\int^{t=t_f}_{t=t_i} f(x^i,\dot x^i)\ dt\, ,\label{eq:nonconstrainedL}
}
with $i =1,\cdots N$. 
The variation gives
\eqn{
\delta S =\l(f_{x^i}-\frac{d}{dt}f_{\dot x^i}\r)\delta x^i +\l[f_{\dot x^i}\delta x^i\r]^{t=t_f}_{t=t_i}\,.
\label{1st-order variation of the action integral of a 1st-order derivative system}
}
Here we used shorthand notations, $f_{x^i}=\frac{\partial f}{\partial x^i}$ and $f_{\dot x^i}=\frac{\partial f}{\partial \dot x^i}$. In order to derive equations of motion and to make $\delta S = 0$, the boundary term is required to vanish. In the usual manipulation, the Dirichlet boundary conditions 
\eqn{
\delta x^i(t_i)=\delta x^i(t_f)=0\quad (i=1,\cdots N)
}
are imposed. Then the variational principle gives equations of motion for the $N$ point particles in the well-posed manner,
\eqn{
0=f_{x^i}-\frac{d}{dt}f_{\dot x^i}\, .
}
Hereinafter in this section, we consider a case where the theory is unconstrained, which is characterized by
\eqn{
\det f_{\dot x^i\dot x^j}\neq 0.
\label{eq:detf}
}
The $N\times N$ matrix $f_{\dot x^i \dot x^j}$ is called the kinetic matrix.

Now, we show that, if \eqref{eq:detf} holds, the minimum number of boundary conditions for a well-posed variational principle is at the least $2N$. This statement is proven by considering the contraposition. Hence, let us assume that the number of boundary conditions can be reduced to $2M$($<2N$). This implies that the boundary term in (\ref{1st-order variation of the action integral of a 1st-order derivative system}) is written with functions $F_a(x^i,\dot x^i)$ and $X^a(x^i)$ where $a$ runs from $1$ to $M(<N)$, that is,
\eqn{
f_{\dot x^i}\delta x^i =F_a \delta X^a.
\label{eq:reduce}
\label{F and X}
}
Therefore, the well-posed variational principle requires the $2M$ boundary conditions to be
\eqn{
\delta X^a(t_i)=\delta X^a(t_f)=0,\quad a=1,\cdots M\, .
}
The relation \eqref{F and X} can be expressed as,
\eqn{
0\=f_{\dot x^i}\delta x^i -F_a \delta X^a\,,\\
\=\l(f_{\dot x^i}-F_a\frac{\partial X^a}{\partial x^i}\r)\delta x^i\,.
\label{eq:delX}
}
Note that \eqref{eq:delX} holds for any $\delta x^i$, which gives 
\eqn{
f_{\dot x^i}-F_a\frac{\partial X^a}{\partial x^i}=0 .
}
Taking the derivative with respect to $\dot x^j$, we obtain,
\eqn{
f_{\dot x^i\dot x^j} =\frac{\partial F_a}{\partial \dot x^i}\frac{\partial X^a}{\partial x^i}\,.
}
Note that $\frac{\partial F_a}{\partial \dot x^i}$ is an $N\times M$ rectangular matrix while $\frac{\partial X^a}{\partial x^i}$ is an $M\times N$ rectangular matrix. Therefore, the determinant of the product of the two rectangular matrices is zero due to Cauchy-Binet's theorem, that is, $\det f_{\dot x^i\dot x^j}=0$. The contraposition means the first statement of paragraph.

Finally, let us confirm that the number of variables for this theory matches with that of the dynamical degrees of freedom, which can be verified in Hamiltonian analysis. The canonical momentum of \eqref{eq:nonconstrainedL} is defined as
\eqn{
p_i :=f_{\dot x^i}\,. 
\label{eq:defpi}
}
Note that if \eqref{eq:detf} holds, the inverse function theorem guarantees that eq.\eqref{eq:defpi} is solved for $\dot x^i$, that is, $\dot x^i = \dot x^i(x^i,p_i)$. Then, the total Hamiltonian is simply written as 
\eqn{
H= p_i\dot x^i(x^i,p_i)-f(x^i,\dot x^i(x^i,p_i))\,.
}
Thus, the phase space spans through all the canonical variables, namely $(p_i,x^i)$ with $i=1,\cdots, N$, which for real space indicates there are $2N$ dynamical degrees of freedom. 

As a conclusion for this section, the number of the boundary conditions for the variational principle coincides with that of the physical degrees of freedom in unconstrained systems, which is twice of the physical degrees of freedom in the terminology of the Hamiltonian analysis.

\section{Procedures for deriving well-posed boundary conditions}
\label{sec:constrained}
In general, boundary conditions for the variational principle in constrained systems are not as simple as their unconstrained counterparts. This is because the number of physical degrees of freedom, which is revealed from performing the Dirac procedure, does not coincide with the actual variables present in the theory. For example, the two point-particles Lagrangians $L =\frac12\dot x^2 +\frac12\dot y^2$ and $\tilde L=\frac12(\dot x+\dot y)^2$ may have the same number of variables but the Dirac procedure reveals that the former has 2 physical degrees of freedom while the latter has only 1. 

Through the equations of motion, boundary conditions determine a solution of a given theory. Too much boundary conditions generically give no solutions, and thus {\it well-posed} boundary conditions are necessary to be imposed. Let us see this point in the viewpoint of the variational principle. We first fix boundary values for some variables. Then, the variation of the action gives equations of motion, if a sufficient number of boundary values are fixed such that the boundary contributions vanish. However, if too much boundary conditions are given by the variational principle, they becomes inconsistent with the equations of motion and the constraint structure that are obtained from the Dirac procedure. Although a final result might sometimes become consistent with too much boundary conditions, it is accidental and does not occur in general situations. Therefore, a sufficient number of boundary conditions for adequate variables are required such that the variational principle holds, but they should not so many as to be inconsistent with the equations of motion and the constraint structure that are revealed by performing the Dirac procedure.

Then, let us introduce a concept of well-posed boundary conditions for the variational principle as those satisfying the following two. 1. The variational principle holds, i.e., the boundary contribution vanishes in the varied action. 2. The minimal possible number of boundary conditions, which turn out to coincide with the number of physical degrees of freedom from performing the Dirac procedure, are imposed while still being consistent with the equations of motion/constraints and all of its solutions. We will show that the following five steps give the well-posed boundary conditions for the variational principle in a given constrained / degenerate Lagrangian system. 
\begin{itemize}
\item[\hypertarget{step1}{Step 1}:] Find the primary constraint(s) and derive the total Hamiltonian
\item[\hypertarget{step2}{Step 2}:] Find all other constraints (secondary, tertiary, etc)
\item[\hypertarget{step3}{Step 3}:] Construct coordinate and momentum variables $(q^{a}_{\perp},p_{a\perp})$ that are orthogonal to the constraints with respect to the Poisson bracket. 
\item[\hypertarget{step4}{Step 4}:] Convert all the Hamiltonian variables back to the Lagrangian variables.
\item[\hypertarget{step5}{Step 5}:] Rewrite the varied boundary terms with respect to the orthogonal variables, up to surface terms, then apply all the constraints.
\end{itemize}
The existence of variables in \hyperlink{step3}{Step 3} is guaranteed by a novel theorem which was proposed by S. Shanmugadhasan \cite{Shanmugadhasan1973,DominiciGomis1980,Dominici1982} and proved by T. Maskawa and H. Nakajima \cite{MaskawaNakajima1976}. This theorem states also that the original coordinate variables of the given system can be transformed in the canonical manner, and guarantees that, in \hyperlink{step4}{Step 4} and \hyperlink{step5}{5}, the boundary term is always rewritten in terms only of the variables. That is, there exists a canonical transformation 
\eqn{
\{q^{i},p_{i}\} \to \{(q^{a}_{\perp},\Xi^{\alpha},\Theta^{\mu} ),(p_{a\perp},\Phi_{\alpha},\Theta_{\mu} )\}
}
where $\Phi_{\alpha}$ and the combination of $\Theta^{\mu}$ and $\Theta_{\mu}$ are the first-class and the second-class constraints, respectively, which are derived from performing the Dirac procedure. The surface term in the variation of an action can be written in
\eqn{
L_{\dot q^i}\delta q^{i} &=& p_{i} \delta q^{i}\nn
&\approx& p_{a\perp} \delta q^{a}_{\perp}+\Phi_{\alpha}\delta \Xi^{\alpha}+\Theta_{\mu}\delta\Theta^{\mu}
+\delta W\nn
&\approx& p_{a\perp} \delta q^{a}_{\perp}
+\delta W,
}
W is a surface term which originates from the canonical transformation and 
$\approx$ is the weak equality in the Dirac procedure. To make the theory well-posed, a counter term $-W$ has to be added to remove the surface term $+W$. Note that terms $\Phi_{\alpha}\delta \Xi^{\alpha}$ and $\Theta_{\mu}\delta\Theta^{\mu}$ disappear not due to taking the variation to be zero, that is, $\delta \Xi^{\alpha}=\delta\Theta^{\mu}=0$, but because the coefficients becomes zero due to the constraints. Therefore, only the boundary condition $\delta q^{a}_{\perp}=0$ is necessary to be imposed and then the variational principle becomes consistent with the equations of motion including the boundary terms with constraints. Then we naturally confirm the coincidence between the number of the physical degrees of freedom, which is none other than the half number of the time-evolutive canonical variables $(q^{a}_{\perp},p_{a\perp})$, and that of the boundary condition(s) for the variational principle.

In order to find pairs of canonical variables $(q^{a}_{\perp},p_{a\perp})$ orthogonal to the constraints in \hyperlink{step3}{Step 3},  the following formula for an arbitrary function $\phi$ gives an insightful clue for cases only with the second-class constraints $\theta_{\bar \mu}=0$,
\eqn{
\phi_{\perp}=\phi-\{\phi,\theta_{\bar \mu}\}(D^{-1})^{\bar \mu \bar \nu}\theta_{\bar \mu}
}
where $D$ is Dirac matrix defined as $D_{\bar \mu \bar \nu}:=\{\theta_{\bar \mu},\theta_{\bar \nu}\}$. We can check that the variable $\phi_{\perp}$ is orthogonal to all the second-class constraints in the sense of the weak equality. Remark, however, that this prescription does not provide canonical pairs but just gives some orthogonal variables. We need further manipulations for the construction of orthogonal canonical pairs, but unfortunately have no universal prescription. However, it is enough to recognize the necessity of additional boundary terms even for first-order derivative theories including the Palatini gravity, as shown in the Sec.\ref{sec:gene}.

\section{Examples of boundary conditions for constrained/degenerate systems}
\label{sec:examples}
In this section, we shall walk-through the five steps in Sec.\ref{sec:constrained} in specific examples.

\subsection{Two-particle system with a constraint}
\label{sec:2pwc}
Let us first start with a simple example of a two-particle system with a constraint. The action considered here is
\eqn{
S_A \=\int^{t_2}_{t_1}L_A dt, \\
L_A \= \frac12\dot x^2 +\frac12 \dot y^2 +\lambda (x+y)\,,\label{eq:L1}
}
which has three variables $(x,y,\lambda)$. Notice that the determinant of the kinetic matrix for $(x,y,\lambda)$ is zero.

The constraint is obtained by varying with respect to $\lambda$,
\eqn{
0=\frac{\delta L_A}{\delta \lambda} =x+y\,. \label{eq:L1constraint}
}
This constraint relates the variables $x$ and $y$, reducing the actual physical degrees of freedom, which we shall look into with more depth later in this subsection.

The variation of the action has boundary terms
\eqn{
\delta S_A=\l[\text{E.o.M.}\r]+\l[\dot x\delta x +\dot y\delta y\r]^{t_2}_{t_1} \,.\label{eq:L1boundary}
}
For a well-posed variational principle, the boundary terms need to disappear, which implies
\eqn{
\l[\dot x\delta x +\dot y\delta y\r]^{t_2}_{t_1}=0 \,.
}
One may be tempted to immediately assume,
\eqn{
0=&\delta x(t_1)&=\delta x(t_2)\,,\\
0=&\delta y(t_1)&=\delta y(t_2)\,.
}
This implies that both $x$ and $y$ are fixed to arbitrary values on the boundary before the variation. However, this fixing in the variational principle generically becomes inconsistent with the constraint given in \eqref{eq:L1constraint}, except accidental cases.

Let us turn to Hamiltonian formalism, that is the Dirac procedure, to see the physical degrees of freedom clearly. The conjugate momenta are
\eqn{
p_x =\dot x\,,\label{eq:L1p1} \qquad
p_y =\dot y\,,\qquad
p_\lambda =0\label{eq:L1p2}\,.
}
We see that there is a primary constraint $0= \Phi :=p_\lambda $. Thus, as \hyperlink{step1}{Step 1}, we write the total Hamiltonian in
\eqn{
H_A=\frac12p_x^2+\frac12p_y^2-\lambda (x+y) +\Lambda \Phi\,.
}
In \hyperlink{step2}{Step 2}, all the constraints for this system are required to be derived. The consistency condition for the primary constraint gives rise to a secondary constraint which then leads to tertiary and quaternary constraints, as, 
\eqn{
\{\Phi,H_A\}=&x+y &:=\Psi\,,\\
\{\Psi,H_A\}=&p_x+p_y &:=2\Xi\,,\\
\{\Xi,H_A\}=&2\lambda&:=2\Theta\,.
}
The constraints ($\Phi,\Psi,\Xi,\Theta$) are all second-class and thus the physical degrees of freedom is $(6-4)/2=1$.

\hyperlink{step3}{Step 3} requires us to find variables orthogonal to the constraints. Such 'proper' coordinate and momentum variables can be taken as,
\eqn{
x_\perp \= x-y\,,\\
p_\perp \=\frac12(p_x-p_y)\,,
}
which are indeed orthogonal to all the constraints. \hyperlink{step4}{Step 4} is to rewrite the Hamiltonian variables into Lagrangian variables, which is achieved with the canonical relations of \eqref{eq:L1p2} as
\eqn{
&& p_\perp=\frac12(\dot x-\dot y)\,, \qquad
\Theta= \lambda\,, \nn 
&& \Xi=\frac12(\dot x+\dot y)\,, \qquad
\Phi=0\,.
}
Thus, in the final \hyperlink{step5}{Step 5}, the boundary term of the varied action \eqref{eq:L1boundary} is rewritten as
\eqn{
\l[\dot x\delta x +\dot y\delta y\r]^{t_2}_{t_1} \=\frac12(\dot x-\dot y)\delta(x-y)+\frac12(\dot x+\dot y)\delta(x+y)\nonumber\\
\=\l[p_\perp \delta x_\perp +\Xi\delta \Psi +\Phi\delta \Theta\r]^{t_2}_{t_1} \,.
\label{ex1surf}
}
Since $\Xi=\Phi=0$ is satisfied for any configuration of solutions including on the boundaries, the only boundary conditions $\delta x_\perp(t_2) =\delta x_\perp(t_1)=0$ are sufficient for the variational principle to be well-posed. 
We emphasize that the last two terms in Eq.\eqref{ex1surf} disappear not because of $\delta \Psi=\delta \Theta=0$ but due to $\Xi=\Phi=0$. 
We do not impose any boundary conditions for $\delta \Psi$ and $\delta \Theta$.
Moreover, any fixing of $x_\perp$ on the boundary is consistent with the equations of motion and constraints.

\subsection{Two-particle degenerate system: First and second-class constraint}
\label{sec:2pwd2nd}
In the previous subsection, a constraint was introduced to a two-particle system through a Lagrangian multiplier. Instead, we investigate a theory where an additional constraint arises due to the structure of the function form of the Lagrangian. This is usually called '{\it degeneracy}'. As a simple example, we introduce an action
\eqn{
S_B\=\int^{t_2}_{t_1}L_B dt, \\
L_B\=\frac12(\dot x+ \dot y)^2 -\frac12m^2(x\pm y)^2\,.\label{eq:L2}
}
For the case with $+$ sign, the variable redefinition from $x+y$ to $q:=x+y$ transform $L_{B}$ into 
\eqn{L'_{B}=\dot{q}^{2}/2-m^{2}q^{2}/2.} 
Thus, this system has obviously one physical degree of freedom, and the corresponding boundary conditions are those for only $q$. In contrast, the case with $-$ sign is nontrivial and requires the 5-step procedure to find well-posed boundary conditions for the well-posed variational principle.

The varied action has the following boundary term,
\eqn{
\delta S_B=\l[\text{E.o.M.}\r]+\l[(\dot x+\dot y)\delta x +(\dot x+\dot y)\delta y\r]^{t_2}_{t_1} \,.\label{eq:L2boundary}
}
For the identification of the well-posed boundary condition, let us proceed to \hyperlink{step1}{Step 1}. The conjugate momenta are
\eqn{
p_x=\dot x+\dot y\,, \qquad
p_y=\dot x+\dot y\,.
}
Clearly, there is a primary constraint $\Phi:=p_x-p_y=0$. The total Hamiltonian is thus,
\eqn{
H_B= \frac12p_x^2+\frac12m^2(x\pm y)^2+\Lambda \Phi\,.
}
As for \hyperlink{step2}{Step 2}, the consistency condition for the primary constraint $\Phi$ gives 
\eqn{
\{\Phi,H_B\}\=-m^2(-1\pm1)(x\pm y)\nonumber\\
\= \l\{\begin{array}{cl}0&\ :+\\2m^2(x-y):=4m^2\Psi &\ :-\\0&\ :m=0
\end{array}
\r.\,.
\nn
}
We see that, for $+$ sign and for $m=0$, the constraint $\Phi$ becomes a first-class constraint, whereas for $-$ sign, $\Phi$ is a second-class constraint.

In all cases, the orthogonal variables, following \hyperlink{step3}{Step 3}, can be taken as,
\eqn{
x_\perp\=x+y\,,\\
p_\perp\=\frac12(p_x+p_y)\,.
}
Note that, for all cases, there is a relation \footnote{Note that at this step 3, any function $f(\Phi)$ satisfying $f(0)=0$ may be possible to be added to $x_\perp$ and $p_\perp$. However, at step 4 this redundancy would disappear.}
\eqn{
p_x\delta x +p_y\delta y=p_\perp\delta x_\perp +\Phi\delta\Psi.
} 
The next step, being \hyperlink{step4}{Step 4}, rewriting the variables and the constraints with respect to Lagrangian variables, they become,
\eqn{
p_{\perp}=\dot x+\dot y\,, \qquad
\Phi=0\,.}
Thus, for the final \hyperlink{step5}{Step 5}, the boundary term of the varied action \eqref{eq:L2boundary} is rewritten as
\eqn{
\l[(\dot x+\dot y)\delta x +(\dot x+\dot y)\delta y\r]^{t_2}_{t_1} =\l[p_\perp \delta x_\perp +\Phi\delta \Psi\r]^{t_2}_{t_1} \,.
}
Since $\Phi=0$ for all configuration of solutions everywhere including the boundaries, $\delta x_\perp(t_2) =\delta x_\perp(t_1)=0$ is sufficient for the variational principle to be well-posed for this theory. 
Moreover, the number of the conditions on each boundary coincides with that of the physical degrees of freedom.

\subsection{One-particle degenerate system with second-order derivatives}
\label{sec:1pwd2d}
The next example is a one-particle theory with second-order derivatives
\eqn{
S_C\=\int^{t_2}_{t_1}L_C dt, \\
L_C\=-\frac12x\ddot x -\frac12m^2 x^2\,.
}
This is none other than a point particle Lagrangian in disguise with a surface term
\eqn{
\tilde L_C =\frac12 \dot x^2 - \frac12m^2 x^2 =L_C +\frac12\frac{d}{dt}(x\dot x).
}
For the point particle Lagrangian $\tilde L_C$, one has to impose only two boundary conditions $\delta x(t_2)=\delta x(t_1)=0$. The varied action of $L_C$, however, becomes
\eqn{
\delta S_C=\l[\text{E.o.M.}\r]+\l[-\frac12x\delta \dot x +\frac12 \dot x\delta x\r]^{t_2}_{t_1} \,.
}
which at first glance needs 4 boundary conditions, $\delta \dot x(t_2)=\delta \dot x(t_1)=\delta x(t_2)=\delta x(t_1)=0$, that is, two more conditions are required in addition to those for $\tilde L_C$.

With the above in mind, let us go through the five steps to see what are the well-posed boundary conditions in the variational principle for the theory of $L_C$. Before that, however, let the order of derivatives lowered
\eqn{
S^\prime_C\=\int^{t_2}_{t_1}L^\prime_Cdt\,,\\
L_C^\prime(x,\dot x, y,\dot y) \=-\frac12x\dot y -\frac12m^2 x^2 +\lambda(\dot x-y)\,,\quad
\label{LagLc'}
}
which is dynamically equivalent to $L_C$ as could be seen by solving $\dot x=y$.\footnote{A general theory is presented in \cite{Pons:1988tj}.} Under variation, this action has the following boundary
\eqn{
\delta S^\prime_C=\l[\text{E.o.M.}\r]+\l[-\frac12x\delta y +\lambda\delta x\r]^{t_2}_{t_1} \,. \label{eq:L3boundary}
}
Now, let us start with \hyperlink{step1}{Step 1}. The conjugate momenta of $L^\prime_C$ are
\eqn{
p_x =\lambda\,, \qquad
p_y =-\frac12 x\,,\qquad
p_\lambda =0\,,
}
which gives three primary constraints 
\eqn{\Phi=p_\lambda, \qquad \Psi =p_x-\lambda,\qquad \Xi=p_y+\frac12x\,.}
Thus the total Hamiltonian becomes
\eqn{
H_C=\frac12m^2 x^2 +\lambda y +\Lambda_1\Phi +\Lambda_2\Psi+\Lambda_3\Xi\,.
}
In \hyperlink{step2}{Step 2}, the evolutions of the primary constraints become
\eqn{
\{\Phi,H_C\}\=-y+\Lambda_2\,,\\
\{\Psi,H_C\}\=-m^2x-\Lambda_1-\frac12\Lambda_3\,,\\
\{\Xi,H_C\}\=-\lambda+\frac12\Lambda_2\,,
}
which leads to a secondary constraint 
\eqn{
\Theta =\frac12y-p_x\,.
}
All the constraints are second-class, which indicates that there are $(6-4)/2=1$ physical degree of freedom, as expected. As \hyperlink{step3}{Step 3}, the variables that are orthogonal to the constraints are taken as
\eqn{
x_\perp=&x-\Xi-\Phi&=\frac12x-p_y-p_\lambda\,,\\
p_\perp=&2p_x+\Theta&=p_x+\frac12y\, .
}
Rewriting the constraints with respect to the Lagrangian variables, in \hyperlink{step4}{Step 4}, we obtain
\eqn{
x_\perp=x\,,\qquad
p_\perp=\lambda+\frac12y\,.
}
For the final \hyperlink{step5}{Step 5}, the boundary term of the varied action \eqref{eq:L3boundary} is rewritten as
\eqn{
\l[-\frac12x\delta y +\lambda\delta x\r]^{t_2}_{t_1} =\l[p_\perp \delta x_\perp -\frac12(x\delta y+y\delta x)\r]^{t_2}_{t_1} \,. \nn
}
The boundary term, at first glance, does not seem to disappear under the proper boundary conditions of $\delta x_\perp(t_2)=\delta x_\perp(t_1)=0$. Therefore, the variational principle with Lagrangian \eqref{LagLc'} does not work well because the boundary terms in the varied action does not vanish. However, since \hyperlink{step5}{Step 5} is achieved up to surface terms, the final two terms can be eliminated by introducing the counter term of \eqn{W=\frac12xy} to the Lagrangian $L^\prime_C$, that is, the Lagrangian is deformed into 
\eqn{
L^{\prime\prime}_C=L^\prime_C +\frac{d}{dt}W.} 
Then the newly introduced Lagrangian $L^{\prime\prime}_C$ indeed has a well-posed variational principle with the boundary condition of $\delta x_\perp(t_2)=\delta x_\perp(t_1)=0$. Explicitly, the new Lagrangian is of the form,
\eqn{
L^{\prime\prime}_C=\frac12\dot xy - \frac12m^2 x^2 +\lambda(\dot x-y)\,.
}
This is none other than the point particle Lagrangian $\tilde L_C =\frac12 \dot x^2 - \frac12m^2 x^2$ once the constraint $y=\dot x$ is substituted. Thus following the steps and obtaining the correct boundary conditions, one may obtain the counter term that is necessary to make the variational principle of the theory well-posed.

\section{General cases}
\label{sec:gene}
In the previous section, we demonstrated how our proposed procedure with the five steps works in specific examples. Here, we will see that it works in the cases where function form is general to some extent.

\subsection{Degenerate two-particle system: Purely kinetic case}
Let us consider a two-particle K-essence type Lagrangian
\eqn{
L_{K} =K(\dot x,\dot \phi)
}
which is assumed to be degenerate i.e. $0= K_{\dot x\dot x}K_{\dot \phi\dot \phi}-K_{\dot x\dot \phi}^2$. We assume that the degree of the degeneracy is unity, that is, the matrix $K_{ij}$ is not null matrix. Without loss of generality, we consider a case where $K_{\dot x\dot x}\neq 0$. Then, in particular, if $K_{\dot x \dot \phi}=0$ then the variables $x$ and $\phi$ are separated like as $L_{K}=f(\dot{x})+c\dot{\phi}$, where $c$ is a constant, and the counter term for the variable $\phi$ obviously becomes $W=-c\phi$. In order to avoid such trivial cases, we further impose a condition: $K_{\dot x \dot \phi}\neq0$. The surface term in the varied Lagrangian is
\eqn{
\delta L_{K}\sim\frac{d}{dt}\l[K_{\dot x} \delta x+K_{\dot \phi}\delta \phi\r]\,.
}
Similar to the examples in the previous section, for a degenerate theory, the imposition of the Dirichlet boundary condition for both $x$ and $\phi$ does not give the well-posed variational principle. 

The conjugate momenta are
\eqn{
p=K_{\dot x}\,,\qquad
\pi=K_{\dot \phi}\,.
}
Since the system is degenerate, there exists a primary constraint, which is written in 
\eqn{
0 = \pi-F(p):=\Phi,
}
at least locally on the $(p,\pi)$-space. Taking a derivative of the momenta with respect to $p$, we obtain
\eqn{
1\=K_{\dot x\dot x}\frac{\partial \dot x}{\partial p}+K_{\dot x\dot \phi}\frac{\partial \dot \phi}{\partial p}\,,\\
\frac{\partial \pi}{\partial p}\=K_{\dot \phi\dot x}\frac{\partial \dot x}{\partial p}+K_{\dot \phi\dot \phi}\frac{\partial \dot \phi}{\partial p}\,.
}
The degeneracy condition gives us a relation between $\pi$ and $p$ 
\eqn{
\frac{\partial \pi}{\partial p} =\frac{K_{\dot x \dot \phi}}{K_{\dot x\dot x}}=\frac{K_{\dot \phi\dot \phi}}{K_{\dot x \dot \phi}}\,.
}
The total Hamiltonian is then
\eqn{
H_{K} = p\dot x +F\dot \phi -K +\Lambda \Phi\,,
}
with $\dot x= \dot x(p)$ and $\dot \phi =\dot \phi(p)$. The consistency condition of the constraint $\Phi$ shows that it is first-class.
\eqn{
\{\Phi,H_K\} = 0\,.
}
Thus there are 2 dynamical degrees of freedom in phase space, that is, 1 physical degree of freedom.

Let us search for a variable $q_\perp =q_{\perp}(x,\phi,p)$ orthogonal to $\Phi$, if it exists. Such a variable $q_\perp$ is required to satisfy the partial differential equation
\eqn{
\{q_\perp,\Phi\}=\frac{\partial q_\perp}{\partial \phi} -\frac{\partial F}{\partial p}\frac{\partial q_\perp}{\partial x}=0\,.
}
By taking a variable $q_\perp$ such that it has a structure
\eqn{
q_\perp =q_\perp\l(\phi+\l(\frac{\partial F}{\partial p}\r)^{-1}x\r)\,,
}
we see that $q_\perp$ is always orthogonal to $\Phi$. The conjugate momentum for this variable is
\eqn{
p_\perp =\frac{1}{2q^\prime_\perp}\l(\pi+F\r)\,,
}
with $q^\prime_\perp =\partial_\phi q_\perp$. Taking $q_\perp =\phi+\l(\frac{\partial F}{\partial p}\r)^{-1}x$ and $p_\perp =\frac12(\pi+F)$, the surface term is rewritten in
\eqn{
&&\hspace{-8mm}
p\delta x+\pi\delta \phi\nonumber\\
\=p_\perp\delta q_\perp+\frac12\Phi\delta\l\{\phi-\l(\frac{\partial F}{\partial p}\r)^{-1}x\r\}\nonumber\\
&&\qquad
+\delta\l[x\l\{p-F\l(\frac{\partial F}{\partial p}\r)^{-1}\r\}\r]\nn
\=K_{\dot \phi}\delta\l\{\phi+\frac{K_{\dot x\dot x}}{K_{\dot x\dot \phi}}x\r\}+\delta\l\{x\l(K_{\dot x} -\frac{K_{\dot x\dot x}}{K_{\dot x\dot \phi}}K_{\dot \phi}\r)\r\}.\nn
\label{bcForK}
}
The variable that should be fixed on the boundary is only $\phi+\frac{K_{\dot x\dot x}}{K_{\dot x\dot \phi}}x$. The counter term of
\eqn{
W:=-x\l(K_{\dot x} -\frac{K_{\dot x\dot x}}{K_{\dot x\dot \phi}}K_{\dot \phi}\r)
}
needs to be added to the Lagrangian $L_{K}$, for well-posedness of the variational principle. Note that, {\it even though the Lagrangian $L_{K}$ has only first-order derivatives, the additional surface terms are required.}

\subsection{Degenerate two-particle system: Separated kinetic and potential terms}
Let us consider a two-particle system where the kinetic and potential terms are separated,
that is, Lagrangian is given as 
\eqn{
L_{V}(x,\phi,\dot{x}, \dot{\phi})= K(\dot{x}, \dot{\phi}) - V(x,\phi).
}
The surface term of this Lagrangian is 
\eqn{
\delta L_{V}\sim\frac{d}{dt}\l[K_{\dot x} \delta x+K_{\dot \phi}\delta \phi\r]\,.
}
We assume that the kinetic term is degenerate but non-zero. Without loss of generality, we can impose conditions for the kinetic term as 
\eqn{
&& M := \begin{pmatrix}
K_{\dot{x}\dot{x}} & K_{\dot{\phi}\dot{x}} \\
K_{\dot{x}\dot{\phi}} & K_{\dot{\phi}\dot{\phi}} \\
\end{pmatrix} ,
\nn
&& \mbox{det} M =0, \label{genedetM}\nn
&& K_{\dot{x}\dot{x}} \neq 0.
}
Conjugate momenta for $x$ and $\phi$ become 
\eqn{
p= K_{\dot{x}}, \qquad \pi= K_{\dot{\phi}} , \label{genedefp}
}
respectively. Since the kinetic matrix is degenerate, $p$ and $\pi$ are related, which gives a primary constraint. We write the primary constraint as 
\eqn{
\pi=F(p),
\label{genecon1}
}
which can be justified by $K_{\dot{x}\dot{x}} \neq 0$. The total Hamiltonian becomes 
\eqn{
&&H_T = H_0 + \lambda (\pi- F), \nn
&&H_0 (p,x,\phi) := H_{\pi=F}. \nn
&&H:=p \dot{x} +\pi \dot{\phi} -K(\dot{x},\dot{\phi}) +V(x,\phi).
}
We define $\Phi$ as
\eqn{
\Phi := \pi -F(p)
}
and then the primary constraint \eqref{genecon1} expressed in
\eqn{
\Phi =0. \label{genePhi}
}
The time evolution of the primary constraint $\Phi$ should be zero, which gives a secondary constraint
\eqn{
0= \dot{\Phi} = \left\{\Phi,H_T\right\}= - V_\phi+ F_p V_x =: G(p,x,\phi).
}

We will show the relation between $F$ and $K$. The primary constraint equation \eqref{genePhi} should be an identity, if we write it in terms of $\dot{x}$ and $\dot{\phi}$. Hence,  we have 
\eqn{
0 = \Phi \left(p=K_{\dot{x}}, \pi=K_{\dot{\phi}}\right) = K_{\dot{\phi}} - F\left(p=K_{\dot{x}}\right).
}
Its variation gives
\eqn{
\left(
K_{\dot{x}\dot{\phi}}- F_p K_{\dot{x}\dot{x}}
\right) \delta \dot{x}
+\left(
K_{\dot{\phi}\dot{\phi}}- F_p K_{\dot{x}\dot{\phi}}
\right) \delta \dot{\phi}=0,
}
that is, 
\eqn{
&&K_{\dot{x}\dot{\phi}}- F_p K_{\dot{x}\dot{x}}=0,
\nn
&&K_{\dot{\phi}\dot{\phi}}- F_p K_{\dot{x}\dot{\phi}}=0.
}
Above two are the same equations because of \eqref{genedetM}, and give
\eqn{
F_p \left(p=K_{\dot{x}}\right) = \frac{K_{\dot{x}\dot{\phi}}}{K_{\dot{x}\dot{x}}}.
}

The Poisson bracket of $\Phi$ and $G$ becomes 
\eqn{
\left\{ \Phi, G \right\} = F_p^2 V_{xx} -2 F_p V_{x\phi}+V_{\phi\phi}.
}
Suppose that the $\left\{ \Phi, G \right\}$ is nonzero for the system has one physical degree of freedom. 

Orthogonal component of a physical variable $f$ to constraints $\Phi$ and $G$ is written as 
\eqn{
f_{\perp} := f + \frac{\{f,\Phi\}}{\{\Phi,G\}} G - \frac{\{f,G\}}{\{\Phi,G\}} \Phi. 
}
The normal component of $p$ is written in 
\eqn{
p_{\perp} = p  - \frac{\{p,G\}}{\{\Phi,G\}} \Phi, 
}
and substituting Eq.~\eqref{genedefp} (note that then $\Phi=0$), we have
\eqn{
p_{\perp} \hat = p  , 
}
where $\hat =$ means that we substitute Eq.~\eqref{genedefp}. This may means $p$ would be a good variable for a momentum of the physical degree of freedom. The normal component of $x$ becomes
\eqn{
x_\perp &=& x + \frac{-F_p G- V_x F_{pp} \Phi}{F_p^2 V_{xx} -2F_p V_{x\phi}+V_{\phi\phi}}
\nn
&\hat = & x + \frac{-F_p G}{F_p^2 V_{xx} -2F_p V_{x\phi}+V_{\phi\phi}}. 
}
Since $x_\perp$ has a term proportional to $G$, the Poisson bracket of $x_\perp$ and $p_\perp$ does not give unity, which means that a pair of $x_\perp$ and $p_\perp$ is not conjugate. It may be interesting to consider a variable $x+ \F (p) \phi$, which is a conjugate partnar of $p$ in $(x,\phi,p,\pi)$ space and corresponds to $\phi+\frac{K_{\dot x\dot x}}{K_{\dot x\dot \phi}}x=\frac{K_{\dot x\dot x}}{K_{\dot x\dot \phi}}(x+\frac{K_{\dot x\dot \phi}}{K_{\dot x\dot x}}\phi)$ in Eq.~\eqref{bcForK} for the case with $V=0$. The normal component of $x+ \F (p) \phi$ is 
\eqn{
&&(x+ \F (p) \phi)_\perp 
\nn{}
&&=x+ \F (p) \phi 
\nn
&&\quad
+ \frac{\left(-F_p + \F \right) G + \left(- V_x F_{pp} + \F_p \phi \{p,G\}\right)\Phi }{F_p^2 V_{xx} -2F_p V_{x\phi}+V_{\phi\phi}}
\nn
&&\hat =  x+ \F (p) \phi + \frac{\left(-F_p + \F \right) G}{F_p^2 V_{xx} -2F_p V_{x\phi}+V_{\phi\phi}}. 
}
Therefore, if $\F = F_p$, the above becomes
\eqn{
(x+ F_p \phi)_\perp 
\hat =  x+ F_p \phi ,
}
which might be a good conjugate pair of $p_\perp$. Actually, the Poisson bracket of $(x+ F_p \phi)_\perp$ and $p_\perp$ gives unity after substitution of Eq.~\eqref{genedefp}.  
Then, the surface term of $\delta L_{V}$ is expected to be written as $p\delta (x+ F_p \phi)$. 

Let us check that the surface term of $\delta L_{V}$ becomes $p\delta (x+ F_p \phi)$. The surface term of $\delta L_{V}$ is
\eqn{
p \delta x + \pi \delta \phi = p \delta x + F(p) \delta \phi.
}
Then, adding the counter term of
\eqn{
W:=\frac{d}{dt} \left(-F \phi +p F_p \phi\right),
}
the original surface term of the varied action becomes
\eqn{
&&K_{\dot x} \delta x+K_{\dot \phi}\delta \phi + \delta(-F  \phi +p F_p \phi)_{p=K_{\dot{x}}} 
\nn
&&\hspace{20mm}
=K_{\dot{x}} \delta \left( x + 
\frac{K_{\dot{x}\dot{\phi}}}{K_{\dot{x}\dot{x}}}\phi \right)
\label{eq:FinalKV}
}
This is consistent with the result obtained in the Hamiltonian analysis. Note, again, that, in order to make the variational principle well-posed, {\it i.e.} to be compatible with the Dirac procedure, { \it the additional surface terms are required, even though the Lagrangian $L_{V}$ has only first-order derivatives.}

\section{Conclusion}
\label{sec:con}
In this paper, the methodology of obtaining the well-posed boundary conditions for the variational principle in constrained / degenerate systems is presented. First, we reviewed the boundary conditions for non-constrained systems. It was noted that the number of variables, that of the physical degrees of freedom, and that of the conditions imposed on each boundary in the variational principle coincide with one another for such a system. On the other hand, it was then pointed out that for constraint systems such relations sometimes do not hold and the number of variables does not match with the number of physical variables. This makes a nonsense of the variational principle, that is, too many constraints on the boundary in the variational principle lead to the inconsistency with the Dirac procedure. Then, we proposed the five steps procedure to give the well-posed boundary conditions for the variational principle in a given constrained/degenerate Lagrangian system. We then proceeded to use these methods to derive the boundary conditions for some specific examples of multi-particle and higher derivative theories. Finally, we have shown that it works in the cases with a general function form and given the explicit formulae for variables to be fixed and surface terms. We emphasize that, even in a theory without higher order derivatives such as Palatini formalism of general relativity, the surface terms are required to be introduced for the well-posed variational principle.

For the future, the extension to field theories can be considered. For instance, general relativity is a degenerate theory in the sense that it has second-derivatives in the action but that its equations of motion remain second order. As it is well-known, in order to obtain the proper (Dirichlet) boundary term, one has to introduce the Gibbons-Hawking-York counter term, which was proposed by \cite{York:1972sj} and then by Gibbons and Hawking in \cite{Gibbons:1976ue}. This still is heavily discussed throughout the ages~\cite{Regge:1974zd,Charap:1982kn,York:1986lje,Brown:1992br,Hawking:1995fd,Parattu:2015gga,Padmanabhan:2014lwa}. The derivation of the Gibbons-Hawking-York term, despite its importance, consists of guessing the functional form of the counter term under the assumptions that the Dirichlet boundary conditions are imposed on boundaries and that the spatial diffeomorphism~\cite{Padmanabhan:2014lwa} is respected.
However, we revealed that the simple Dirichlet boundary conditions are not necessarily adequate in degenerate systems. It implies that the systematic derivations, rather than heuristic ones, of counter terms including the well-known Gibbons-Hawking-York counter term should be performed under the appropriate boundary conditions derived by the five-step procedures in \S\ref{sec:constrained}. Although the simple Dirichlet boundary condition miraculously works well for general relativity, it may not for gravitational theories with more constraints, such as Palatini theories of gravity~\cite{Saez-ChillonGomez:2020afj,Gomez:2021roj}. These arguments will be discussed elsewhere in future.
\\
\\
\section*{Acknowledgments}
We would like to deeply thank Mu-In Park for his insightful and sharp question that initiated this work. K.S. would also like to thank Katsuki Aoki, Shoichiro Miyashita, and Alexander Vikman for their fruitful and helpful discussions. K.I. is supported by JSPS Grant-in-Aid for Scientific Research Numbers JP17H01091, JP21H05182, JP21H05189, JP20H01902, and JSPS Bilateral Joint Research Projects (JSPS-DST collaboration) for Scientific Research Number JPJSBP120227705. K.S. was supported by JSPS KAKENHI Grant Number JP20J12585 during the initial stages of this work. K.T. is supported by Tokyo Tech Fund Hidetoshi Kusama Scholarship. M.Y. is supported by IBS under the project code, IBS-R018-D3, and by JSPS Grant-in-Aid for Scientific Research Number JP21H01080. 
\bibliography{bcforcsbib}
\bibliographystyle{unsrt}
\end{document}